\documentclass{article}

\usepackage[final]{nips_2017}

\usepackage[utf8]{inputenc} 
\usepackage[T1]{fontenc}    
\usepackage{hyperref}       
\usepackage{url}            
\usepackage{booktabs}       
\usepackage{amsfonts}       
\usepackage{nicefrac}       
\usepackage{microtype}      
\usepackage{graphicx}

\title{A Cascade Architecture for Keyword Spotting on Mobile Devices}

\author{
  Alexander Gruenstein, Raziel Alvarez, Chris Thornton, Mohammadali Ghodrat \\
  Google Inc. \\
  1600 Amphitheatre Parkway, Mountain View, CA 94043 \\
  \texttt{\{alexgru,raziel,thorntonc,ghodrat\}@google.com}
}

\begin{document}

\maketitle

\begin{abstract}
  We present a cascade architecture for keyword spotting with
  speaker verification on mobile devices. By pairing a small
  computational footprint with specialized digital signal processing (DSP)
  chips, we are able to achieve low power consumption while
  continuously listening for a keyword.
\end{abstract}

\section{Introduction}
Voice assistants have become prevalent in the last few years. A common
way to initiate a conversation with your assistant is via a keyword,
such as ``Ok Google'', ``Alexa'', or ``Hey Siri''. Thus, it is
important for the keyword to be available in as many devices as
possible, many of them battery powered or with restricted
computational capacity.

Keyword detection is like searching for a needle in a haystack: the
detector must listen to continuously streaming audio, ignoring nearly
all of it, yet still triggering correctly and instantly. On a mobile
device, this is particularly challenging when considering that typical
mobile devices (e.g. smartphones) have batteries with capacities
between 1,000mAh and 2,400mAh, and thus in the best case our entire
system must consume less than 1mA to consume less than 1\% of the
battery per day.

We present a cascade architecture for keyword spotting that can meet
these power requirements, while achieving high accuracy. The first
stage is comprised of a very small and power efficient detector that
executes on a DSP. Upon trigger, it delegates the final detection
decision to a second, much larger and more accurate detector, that
executes on the device's main application processor (AP).

The following section gives an overview of the system. Subsequent
sections describe our quantization approach, as well as a final stage
of the pipeline that runs on the server. Finally we present accuracy results
of the cascade system.

\section{Architecture}
\begin{figure}
  \caption{Cascade keyword spotting architecture}

  \label{diagram}
  \begin{center}
    \includegraphics[height=0.7in]{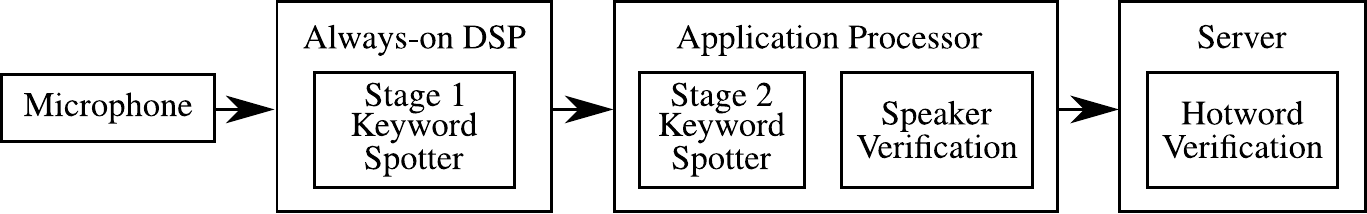}
  \end{center}
\end{figure}

Our system, shown in figure~\ref{diagram} is a cascade of two keyword
detectors that trade-off computation and memory footprint (and thus
power consumption), with accuracy. The first stage is very small and
executes continuously, buffering enough audio to safely fit the
keyword (typically 2 seconds). Upon detection it passes the audio
buffer to the second stage, and continues to stream the audio that
follows. The second stage uses uses a larger and more accurate model
to make the final decision. The second stage also supports speaker
verification, allowing only known speakers to trigger the keyword.

This cascade approach is advantageous regardless of where the
execution takes place, however when running continuously it is
impossible to meet the previously described power restrictions using a
standard mobile AP. Thus, we utilize power-isolated memory and
processor, typically in the form of a DSP, to execute the continuous
first stage. The second stage, with more relaxed power constraints, is
both able to and required to execute a larger and better quality
detection model to overcome the leniency of the first stage. However,
computation is still quite important, as the second stage must be able
to process the buffered audio coming from the first stage as fast as
possible to minimize detection latency.  Given this different set of
requirements, we execute the second stage on the AP.

\subsection{Anatomy of a keyword spotter}
Our detectors receive the raw audio as input and produce a yes/no decision.
Internally they are composed by three main components: 1) a signal
processing frontend 2) a neural network acoustic encoder and 3) a decoder.

\textbf{Frontend}
The frontend receives the audio signal and produces features for the encoder
neural network. Its execution involves some typical signal processing
algorithms, like spectral subtraction, but ultimately produces a log-mel
filterbank (the log of the triangular mel filters applied to the power
spectra). We typically use either 32 or 40 filterbank channels per 25ms frame.

\textbf{Neural network acoustic encoder}
The encoder receives the filterbank channels from one or more stacked frames (depending
on the neural network topology), and produces posterior probabilities for
acoustic units from the keyword (e.g. phones or syllables). This component is
what dominates both the computation as well as the ultimate quality of the
system, and consequently it is where a lot of research has been focused on,
iterating over several topologies and training setups \cite{hotwordv1,localconn,cnn}
until arriving to the rank constrained topology proposed in \cite{svdf}.

\textbf{Decoder}
The decoder receives the $M$ posteriors for the acoustic units of the keyword,
$\mathbf{y} = \left\{ y_1, y_2, \cdots, y_M\right\}$, and produces a score
$h(\mathbf{y})$ between 0 and 1. The score is computed by smoothing the
posterior values, $s_t(y_i)$, by averaging them over a sliding
window of the previous L frames with respect to the current $t$.
The score is then defined as the largest product of the smoothed posteriors in
the input sliding window, subject to the constraint that the individual units
‘fire’ in the same order as specified in the keyword, as described in \cite{agc}
by the following formulation:

\begin{equation}
s_t(y_i) = \frac{1}{L} \sum_{j=t - L + 1}^{t} y_{i,j}; \quad h(\mathbf{y}) = \left[ \max_{1 \leq t_1 \leq \cdots \leq t_M \leq T_s} \prod_{i=1}^M s_{t_i}(y_i) \right]^\frac{1}{M}
\label{eq:keyword-score}
\end{equation}

\subsection{Speaker verification}
On mobile devices, speaker verification is used so that the device
responds only to its owner. It works by "enrolling" the speaker on N
(typically 3) utterances of the keyword, creating a unique
signature. At test time, the utterance, segmented by the second
stage's keyword detector, is processed by the speaker verifier to
generate another signature and compare it with that stored on-device,
as described in \cite{siddnn}. The comparison, typically a cosine
distance between the 2 signatures, will produce a score on which a
threshold can be selected to determine verification status.

The speaker verifier is a neural network, typically an LSTM as in \cite{sidlstm},
which processes frames generated by the frontend, segmented by the second stage
keyword detector. This neural network emits the speaker signature in the form
of a vector of numbers.

\subsection{Operating point selection}
The accuracy of an individual detector, and of the cascade, is
measured as a tradeoff between false accept rate (FAR) -- triggering
for audio not containing the keyword -- and false reject rate (FRR) --
failing to trigger for audio containing the keyword. For the cascade
to work properly, the first and second stage keyword detectors must be
trained on the same data and have similar FRR. This way, instead of being
additive, their combined FRR is very close to that of the second stage
alone.

Since the first stage model is significantly smaller, it follows that
its FAR must be higher than that of the first stage. The FAR is still
quite important, as each run of the second stage will briefly draw 1
to 2 orders of magnitude more current. If the FAR of the first stage
is too high, the combined power consumption will be significant.  We
are typically able to tune the first stage to only wake up a handful
of times per hour, even when exposed continuously to speech.

Speaker verifications acts as a third filter. It contributes both to
increasing the FRR (when it incorrectly rejects a keyword spoken by
the enrolled speaker) and to decreasing the FAR. We find that the
speaker verification can reduce the overall FAR by a factor of 5 to 10
while adding less than 1\% absolute additional FRR, since keyword
spotting false alarms from other people speaking, television shows,
radio, etc are almost always rejected.

\subsection{Considerations of the DSP implementation}
\textbf{Memory}
DSPs come in a variety of forms and specifications: they can be integrated as
part of a system on chip (SoC), or connected externally. However a common
trait is that in order to achieve power efficiency and to reduce cost,
DSPs typically have small amounts of memory (e.g. 128kB is typical).

The available memory is split between code, working-buffers and any
loaded artifacts (e.g. neural network models). In addition, to achieve
low latency the model and code must be kept resident in memory at all
times. This means, that code and model size are very important. Our
system's program memory takes up approximately 25kB, plus another 12kB
for things like fast Fourier transform (FFT) twiddle tables, leaving
the rest of available memory to fit the model and the audio buffer.
2 seconds of 16bit mono PCM audio at a 16khz sample rate consumes 64kB.
That means that given the aforementioned 128kB restriction, we
have set the footprint of our first stage keyword model to 13kB.

\textbf{Precision}
DSP platforms have their own instruction sets, thus each requires
specialized optimization work. Moreover, as computation can vary from 16, 24,
or 32 bits, and platform-specific optimization such as FFT and matrix
multiplication can produce somewhat different results, we cannot guarantee that
the computation done on each platform is bit identical.

This makes scaling to supporting multiple DSP platforms challenging,
as we need to be able to guarantee correctness per platform. Moreover,
differences in the signal processing frontend can sometimes be significant
enough that training platform-specific neural networks becomes desirable.

To accommodate this, we use emulation libraries to produce bit
identical behavior using standard C code, emulating the specifics of
each platform. This way we can still perform large scale neural
network training and evaluation, resulting in per-platform neural
network models.

\section{Quantization}
In order to reduce the memory footprint of the neural network models,
take advantage of the available fast integer operations (e.g. matrix
multiplications), and cover the broadest range of DSPs, we transform
the inputs, as well as the trained parameters of the neural network models
from their original floating point representation into an integer based one
(8 bit to be precise).

The parameter quantization itself can be defined as using a uniform linear
quantizer that assumes a uniform distribution of the values within a given
range. First, we find the minimum and maximum values of the original
parameters. We then use a simple mapping formula which determines a scaling
factor that when multiplied by the parameters spreads the values evenly in the
smaller precision scale, thus obtaining a quantized version of the original
parameters. We have slightly different approaches for executing inferences on
the quantized models: as previously mentioned, on a DSP all execution is fixed
point, whereas in the CPU we take advantage of floating point
accumulators as proposed in \cite{quant}.

\section{Server-side validation}
A third, optional, stage occurs on the sever side as part of the full
speech recognizer, which can further reduce the false accept
rate~\cite{serverhotword}. In addition, the keyword spotter produces
alignment information, which can improve start-of-speech
detection. Because of this, and coarticulation effects, the overall
word error rate can be improved on the query.

\section{Results}
Table~\ref{resultstable} shows the accuracy of the cascaded
keyword spotter as a function of the operating point of the first
stage. The FRR approaches that of the stage 2 model, shown on
the first line, as the stage 1 model becomes more lenient. FAR is
measured as false alarms per hour on a recording of 924 hours of
television background noise. FRR is measured on 65,581 recordings of
US English speakers saying a keyword (redacted in anonymized
version of paper).



\begin{table}[t]
  \label{resultstable}
  \caption{Cascade operating point as a function of the stage 1
    operating point.}
  \center
  \begin{tabular}{cccc}
    \toprule
    \textbf{Stage 1 FA/hr} & \textbf{Stage 1 FRR} & \textbf{Cascade FA/hr} & \textbf{Cascade FRR} \\
    \midrule
    None & None & .03 & 3.1\% \\
    0.5 & 4.1\% & 0.006 & 5.6\%  \\
    0.8 & 3.4\% & 0.006 & 5.1\%  \\
    5.0 & 1.6\% & 0.02 & 3.8\% \\
    10.0 & 1.2\% & 0.02 & 3.5\% \\
    \bottomrule
  \end{tabular}
\end{table}






\small
\bibliographystyle{plain}
\bibliography{wakeword}

\end{document}